\documentclass{emulateapj}
\def\({\left(}
\def\){\right)}
\def\[{\left[}
\def\]{\right]}

\usepackage{graphicx}
\usepackage{natbib}
\usepackage[usenames,dvips]{color}

\begin{document}
\title{Constraints on the Viscosity and Magnetic Field in Hot Accretion Flows around Black Holes}
\author{B. F. Liu\altaffilmark{1}
\and Ronald E. Taam\altaffilmark{2,3}}
\altaffiltext{1}{National Astronomical Observatories, Chinese Academy of Sciences, 
20A Datun Road,  Chaoyang District, Beijing 100012, China; bfliu@nao.cas.cn}
\altaffiltext{2}{Academia Sinica Institute of Astronomy and Astrophysics-TIARA, P.O. Box 23-141, 
Taipei, 10617 Taiwan; r-taam@northwestern.edu}
\altaffiltext{3}{Department of Physics and Astronomy, Northwestern University,  2131 Tech Drive, 
Evanston, IL 60208, USA}

\begin{abstract}
The magnitude of the viscosity and magnetic field parameters in hot accretion flows is investigated in 
low luminosity active galactic nuclei (LLAGNs).  Theoretical studies show that a geometrically thin, 
optically thick disk is truncated at mass accretion rates less than a critical value by mass 
evaporated vertically from the disk to the corona, 
with the truncated region replaced by an advection dominated accretion flow (ADAF). The critical 
accretion rate for such a truncation is a function of the viscosity and magnetic field. Observations 
of X-ray photon indices and spectral fits of a number of LLAGNs published in the literature provide 
an estimate of the critical rate of mass accretion and the truncation radius respectively. By 
comparing the observational results with 
theoretical predictions, the viscosity and magnetic field parameters in the hot accretion flow region 
are estimated. Specifically, the mass accretion rates inferred in different sources constrain the 
viscosity parameter, whereas the truncation radii of the disk, as inferred from spectral fits, further 
constrain the magnetic field parameter. It is found that the value of the viscosity parameter in the 
corona/ADAF ranges from 0.17 to 0.5, with values clustered about 0.2-0.3. 
Magnetic pressure is required by the relatively small truncation 
radii for some LLAGNs and is found to be as high as its equipartition value with the gas pressure. The 
inferred values of the viscosity parameter are in agreement with those obtained from the observations of 
non-stationary accretion in stellar mass black hole X-ray transients.  This consistency provides support 
for the paradigm that a geometrically thin disk is truncated by means of a mass evaporation process from 
the disk to the corona at low mass accretion rates.
\end{abstract}

\section{Introduction}

The $\alpha$ viscosity prescription (Shakura \& Sunyaev 1973) has proved to be a particularly valuable 
framework for accretion disk models, having been used successfully in the interpretation of observations 
of diverse phenomena ranging from dwarf novae and black hole X-ray binaries (BHXRBs) on the small scale 
to active galactic nuclei (AGN) on the large scale. The physical mechanism of the viscosity, however, 
remains to be completely understood. Hydrodynamic as well as magnetohydrodynamic (MHD)
turbulence have been proposed as the 
main sources (e.g. Balbus \& Hawley 1991; Kato 1994), but the deduced viscosity parameter is 
smaller by an order of magnitude in comparison to the values required by observations as inferred from 
time-dependent accretion in X-ray transients (e.g. Smak 1999; Suleimanov et al. 2008; King et al. 2007).  

It is generally accepted that the magnetic field is an important ingredient in the description of 
accretion flows and their emission.  In particular, it is likely to be responsible for the accretion disk 
viscosity as suggested in the early work by Balbus \& Hawley (1991). The emission due to synchrotron radiation 
and self-Compton scattering in advection dominated accretion flows (ADAFs) is sensitive to the strength 
of the magnetic fields and the hard X-ray emission in some objects can be due to magnetic flaring 
activity. Finally, a large scale magnetic field, as required for the formation of jets, can originate from 
small scale magnetic fields produced in the accretion flow by dynamo action.  Although magnetic fields 
play a key role in the above, the strength of such magnetic fields in these flows remains unknown.   
 
The mode of the accretion flow is generally believed to depend on the Eddington-scaled accretion rate. At 
high mass accretion rates, as revealed in the high/soft state of BHXRBs and luminous AGNs, the accretion 
is thought to occur via a geometrically thin, optically thick accretion disk as developed by Shakura \& 
Sunyaev (1973).  However, at low accretion rates in the low/hard state of BHXRBs and low luminosity AGNs 
(LLAGNs), the accretion occurs via a geometrically thick, optically thin ADAF (see Narayan \& Yi 1994, 
1995a, 1995b; Wandel \& Liang 1991), connecting to an outer geometrically thin disk. The transition between these modes can 
be due to a thermal instability (Takeuchi \& Mineshige 1998, Gu \& Lu 2000, Lu et al. 2004), a 
result of radial conduction (Honma 1996, Manmoto \& Kato 2000, Gracia et al 2003) or a disk evaporation 
process (Meyer, et al.  2000a, 2000b, R\`o\.za\`nska \& Czerny 2000a,b, Spruit \& Deufel 
2002, and Dullemond \& Spruit 2005).  In the context of these studies, the disk evaporation model is 
the most promising, which has been investigated to elucidate the disk truncation and spectral 
state transition behavior as a function of accretion rate in both BHXRBs and AGNs (e.g. Liu et al. 1999; 
2002; 2009; Meyer et al. 2000a,2000b; 2007; Taam et al. 2012).

Here, we adopt the disk evaporation model as the mechanism of disk truncation and spectral state 
transition. In this model, the accretion rate characterizing the transition as well as the truncation 
radius is a function of the viscosity and magnetic field parameters.  Detailed spectral fitting to 
the transition or low/hard state spectrum can, in principle, yield these properties. An estimate of 
the viscosity and magnetic field parameters follows from comparing the model prediction with 
observations. From statistical studies based on observed spectra a transition of the spectral 
energy distribution at a determined Eddington ratio can be used to constrain the average value of the 
viscosity parameter in the accretion flows around AGNs. In the next section, we provide a description 
of disk truncation and spectral transition within the context of the mass evaporation model. Constraints 
on the viscosity and magnetic field parameters as obtained from the observations of AGN are provided 
in \S 3.  We discuss our results and conclude in the last section. 
  
\section{Disk Truncation and Spectral Transition as a Consequence of Disk Corona Interaction}
In the disk corona model (for dwarf novae see Meyer \& Meyer-Hofmeister 1994; for black holes see 
Meyer et al. 2000a and Liu et al. 2002), a hot optically thin corona is presumed to 
lie above and below a geometrically thin standard disk, which could be formed by processes similar to 
those operating in the surface of the Sun, or by a thermal instability in the uppermost layers of the 
disk (e.g., Shaviv \& Wehrse 1986).  Both the disk and corona are individually powered by the release 
of gravitational energy associated with the accretion of matter affected through viscous stresses.  In 
the corona, the viscous heat is partially transferred to the electrons by means of Coulomb collisions 
and partially advected radially inward.  The energy gained by the electrons is mainly conducted into 
the lower, cooler, and denser layer and radiated away in the chromosphere.  If the density in the lower 
corona is too low to efficiently radiate this energy, which is the case for a steady accretion corona, a 
suitable amount of cool matter is heated and evaporated into the corona.  The mass accretion in the 
corona is maintained by a steady evaporation flow from the underlying cool disk.

In the accretion disk, gas is partially evaporated into the corona on the way to the accreting black 
hole. This diverts a part of the disk accretion flow into the corona. If the mass supply rate to the 
outer disk is too low, the disk accretion flow can be completely evaporated into the corona at some 
distance from the black hole, interior to which there exists only a hot accretion flow.  This results in 
disk truncation and a change of accretion flow from two-phase (cold disk + hot corona) accretion in 
the outer region to an ADAF in the inner region.  On the other hand, if the mass supply rate to the 
disk is high, the evaporation is limited by efficient Compton cooling of the corona due to the 
strong soft photon field originating from the inner disk. In this case, only a small fraction of the 
mass flow in the disk is evaporated into the corona. Hence, a geometrically thin disk extends to the 
innermost stable circular orbit (ISCO) and the corona is quite weak.  Previous investigations of the 
disk corona interaction (e.g. Meyer et al. 2000a; Liu et al. 2002) reveal that there exists a maximal 
evaporation rate, above which the mode of accretion is dominated by a geometrically thin, optically 
thick accretion disk, and below which the accretion is dominated by an ADAF connected to a truncated 
outer disk with overlying corona.  The maximal evaporation rate, thus, represents the accretion 
rate at the spectral state transition, and the corresponding truncation radius is the minimal radius 
of disk truncation before transit to a soft state.  

The mass evaporation is a consequence of hydrodynamic equilibrium. It is calculated by solving a set of 
simplified  differential equations including the continuity equation, momentum equation, and energy equations 
supplemented by an equation of state (for details see Liu \& Taam 2009), which are listed as follows:

\begin{equation}\label{e:EOS}
P={\Re \rho \over 2\mu} (T_{i}+T_{e}),
\end{equation}

\begin{equation}\label{e:continuity}
\centering
 {d\over dz}(\rho v_z)={2\over R}\rho v_R -{2z\over
R^2+z^2}\rho v_z,
\end{equation}

\begin{equation}\label{e:mdot}
\rho v_z {dv_z\over dz}=-{dP\over dz}-\rho {GMz\over
(R^2+z^2)^{3/2}},
\end{equation}

\begin{equation}\label{e:energy-i}
\begin{array}{l}
{d\over dz}\left\{\rho_i v_z \left[{v^2\over 2}+{\gamma\over
\gamma-1}{P_i\over \rho_i}-{GM\over (R^2+z^2)^{1\over
2}}\right]\right\}\\
={3\over 2}\alpha P\Omega-q_{ie}\\
+{3\over R}\rho_i v_R
\left[{v^2\over 2}+{\gamma\over \gamma-1}{P_i\over \rho_i}-{GM\over (R^2+z^2)^{1\over 2}}\right]\\
-{2z\over {R^2+z^2}}\left\{\rho_i v_z \left[{v^2\over 2}+{\gamma\over
\gamma-1}{P_i\over \rho_i}-{GM\over (R^2+z^2)^{1\over 2}}\right]\right\},
\end{array}
\end{equation}

\begin{equation}\label{e:energy-t}
\begin{array}{l}
{\frac{d}{dz}\left\{\rho {v}_z\left[{v^2\over
2}+{\gamma\over\gamma-1}{P\over\rho}
-{GM\over\left(R^2+z^2\right)^{1/2}}\right]
 + F_c \right\}}\\
=\frac{3}{2}\alpha P{\mit\Omega}-n_{e}n_{i}L(T_e)-q_{\rm Comp}\\
+{3\over R}\rho v_R \left[{v^2\over
2}+{\gamma\over\gamma-1}{P\over\rho}
-{GM\over\left(R^2+z^2\right]^{1/2}}\right]\\
-{2z\over R^2+z^2}\left\{\rho v_z\left[{v^2\over
2}+{\gamma\over\gamma-1}{P\over\rho}-
{GM\over\left(R^2+z^2\right)^{1/2}}\right]
+F_c\right\},
\end{array}
\end{equation}
where Eq.(\ref{e:energy-i}) is the energy equation for the ions, in which $q_{ie}$ is the 
energy exchange rate between the electrons and the ions,
\begin{equation}
{q_{ie}}={\bigg({2\over \pi}\bigg)}^{1\over 2}{3\over 2}{m_e\over
m_p}{\ln\Lambda}{\sigma_T c n_e n_i}(\kappa T_i-\kappa T_e)
{{1+{T_*}^{1\over 2}}\over {{T_*}^{3\over 2}}}
\end{equation}
with
\begin{equation}
T_*={{\kappa T_e}\over{m_e c^2}}\bigg(1+{m_e\over m_p}{T_i\over
T_e}\bigg).
\end{equation}

Eq.(\ref{e:energy-t}) is the  energy equation for both ions and electrons, where $n_{e}n_{i}L(T_e)$ is the bremsstrahlung cooling rate and  ${q_{\rm {Comp}}}$  is the Compton cooling rate,
\begin{equation}\label{e:comp}
 {q_{\rm Comp}}={4\kappa T_e\over m_e
c^2}n_e\sigma_T c u,
\end{equation}
with $u$ the energy density of the soft photon field.  The thermal conduction flux,
$F_c$, is given by (Spitzer 1962)
\begin{equation}\label{e:fc}
F_c=-\kappa_0T_e^{5/2}{dT_e\over dz}
\end{equation}
with $\kappa_0 = 10^{-6}{\rm erg\,s^{-1}cm^{-1}K^{-7/2}}$ for a fully ionized plasma.

All parameters in the above equations are in cgs units and are defined as follows.
Specifically, $P,\rho,T_i$ and $T_e$ are the pressure, density, ion temperature and
electron temperature respectively.  The vertical and radial speed of the mass flow are
denoted by $v_z$ and $v_R$. The other quantities are as follows:  $G$ is the
gravitational constant, $M$ the mass of the accreting black hole, $m_p$ and $m_e$ are respectively
the mass of the proton and the electron, $\kappa$ is the Boltzmann constant, $c$ the speed of light,
$a$ the radiation constant, $\sigma$ Stefan-Boltzmann constant, $\sigma_T$ the Thomson scattering
cross section, $\gamma=5/3$ is the ratio of specific heats, $\mu=0.62$ is the mean molecular weight and $\ln\Lambda=20$ is the Coulomb logarithm.

The evaporation rate (as a function  of $\rho$ and  $v_z$ at the lower 
boundary) depends on the 
heating and cooling processes (see also  Begelman \& McKee 1990; Meyer \& Meyer-Hofmeister 1994; 
R\`o\.za\`nska \& Czerny 2000a;b).   In the corona, viscously released accretion energy  is the  source for 
heating. Cooling processes include downward conduction, energy flux taken by the accreting flow, 
radiative cooling through Bremsstrahlung, Synchrotron and Compton scattering.  At low accretion 
rates/hard states, as considered in this work,  radiative cooling in the corona is  
negligible  compared to the radial advection and vertical conduction.  Only in the transition layer is 
the Bremsstrahlung radiation important.   This implies that the evaporation characteristics are only 
weakly dependent on the radiation in the low hard states, which has been confirmed by numerical calculations 
(e.g. Meyer et al. 2000a; Liu et al. 2002). 

The maximal mass evaporation rate in the low state depends on the viscosity 
parameter, $\alpha$.  An increase in $\alpha$ leads to an increase in heating,  which is partially transferred 
to the electrons through Coulomb collisions and conducted down to the transition layer, resulting in 
an increase in the mass evaporation rate.  This effect is more important in the inner region of the corona 
since the Coulomb collisions in its outer region are very inefficient due to the low densities.
Specifically, the maximum evaporation rate and its corresponding 
truncation radius vary with $\alpha$ approximately as (Qiao \& Liu  2009)
\begin{equation}\label{tr-alpha}
\dot m_{\rm max}\approx 0.38\alpha^{2.34}\
{\rm and}\  r_{\rm min}\approx 18.80\alpha^{-2.00}.
\end{equation}
Here, the evaporation rate is expressed in terms of the Eddington mass accretion rate and the radius 
in terms of Schwarzschild radii, $R_S = 2GM/c^2$, where M is the mass of the black hole. 

The effect of magnetic fields is a competition between its tendency to increase the evaporation as a result of energy balance and to decrease the evaporation as a result of pressure balance. The additional pressure contributed by the magnetic fields results in a greater heating via the shear stress. This effect is 
similar to an increase in $\alpha$ and leads to an increase of evaporation rate in the inner region 
with little effect in the outer region. On the other hand, the additional pressure contribution 
inhibits the evaporation at all distances as a result of force balance. The combined effect of the magnetic field leads to little change in the 
value of the  maximal evaporation rate, but does lead to an inward  shift of the maximal 
evaporation region to a smaller distance.  As shown from our numerical calculations, the maximum 
evaporation rate varies only slightly with the magnetic field strength, parameterized by the ratio of gas 
pressure to the total pressure, $\beta$, though the truncation of the disk occurs at much smaller 
radii for smaller $\beta$  (for details see Qian, Liu, \& Wu 2007). The dependence of the maximum evaporation rate and the corresponding 
truncation radius (which is the minimal truncation radius) on the magnetic field parameter for 
$\alpha=0.3$ can be approximated by  
\begin{equation}\label{tr-beta}
\dot m_{\rm max}\approx 0.026\beta^{-0.41}\  {\rm and} \  r_{\rm min}\approx 209\beta^{4.97}.
\end{equation}

Taking into account these two effects, we find that the maximal evaporation rate and corresponding 
truncation radius is dependent on the viscosity and magnetic parameters as (see Taam et al. 2012),
\begin{eqnarray}
\dot m_{\rm max} \approx 0.38\alpha^{2.34} \beta ^{-0.41} \label{mdot-max}\\
r_{\rm min} \approx 18.80\alpha^{-2.00}\beta^{4.97}\label{tr-min}.
\end{eqnarray}

\begin{figure}
\plottwo{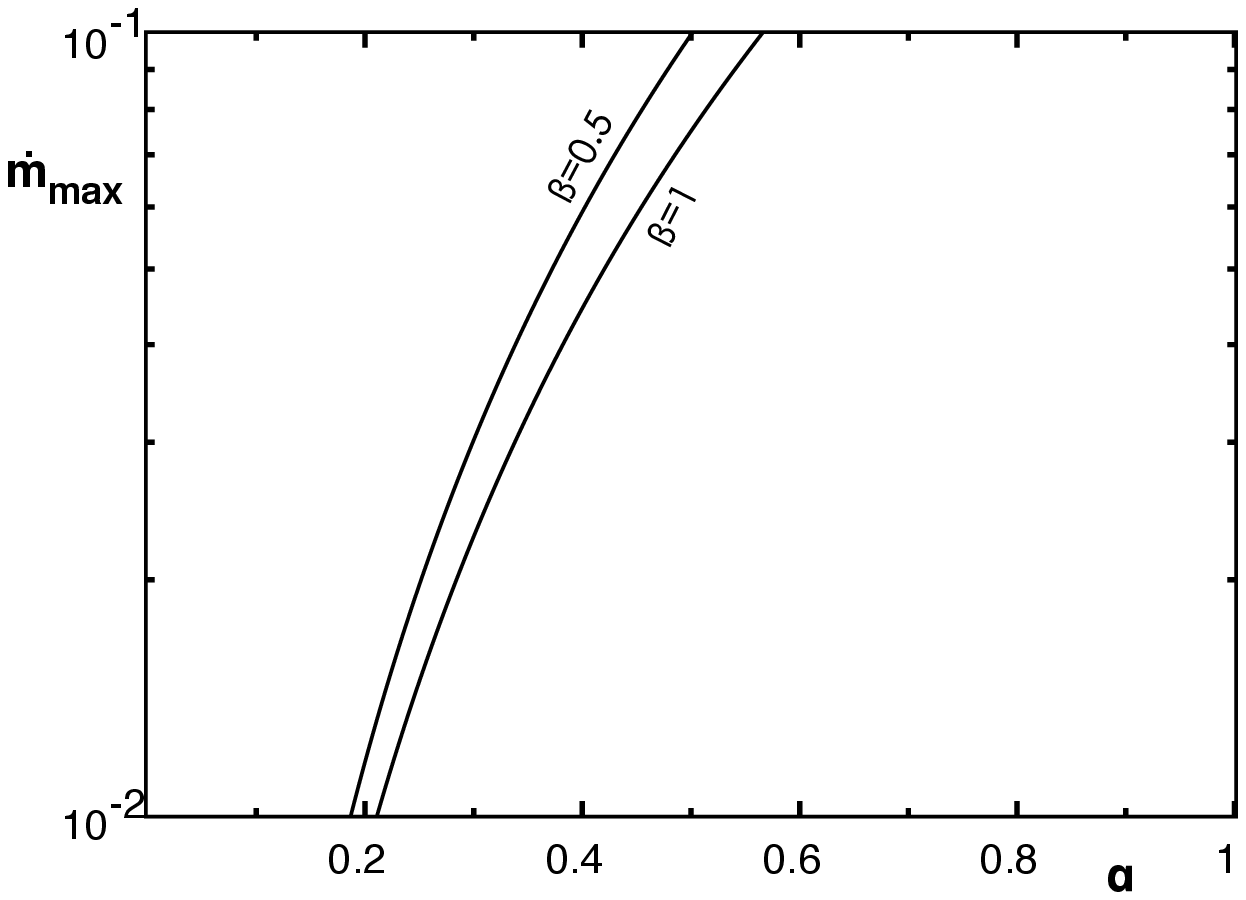}{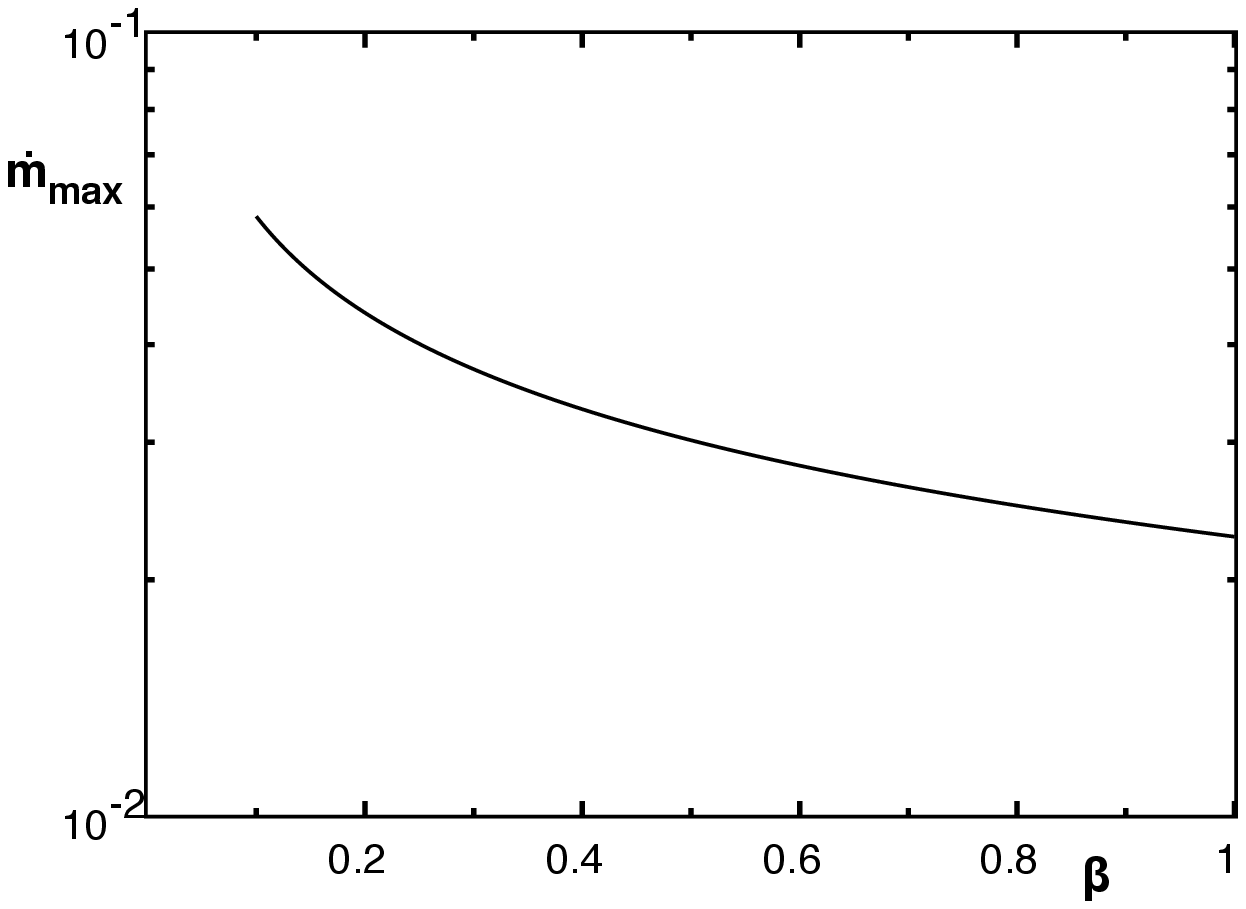}
\caption{\label{f:mdot_trs-alpha} The variation of the critical accretion rate at the spectral state 
transition with respect to the viscosity parameter $\alpha$ (left panel) and magnetic field parameter 
$\beta$ (right panel) as predicted by the disk evaporation model. The transition from the hard state 
to soft state takes place at lower accretion rates for a lower value of $\alpha$. The effect of the magnetic 
field on the transition rate is limited to within the two solid curves in the left panel, which can also be seen from the right panel for a given viscosity parameter, $\alpha=0.3$.}
\end{figure}

\begin{figure} 
\plottwo{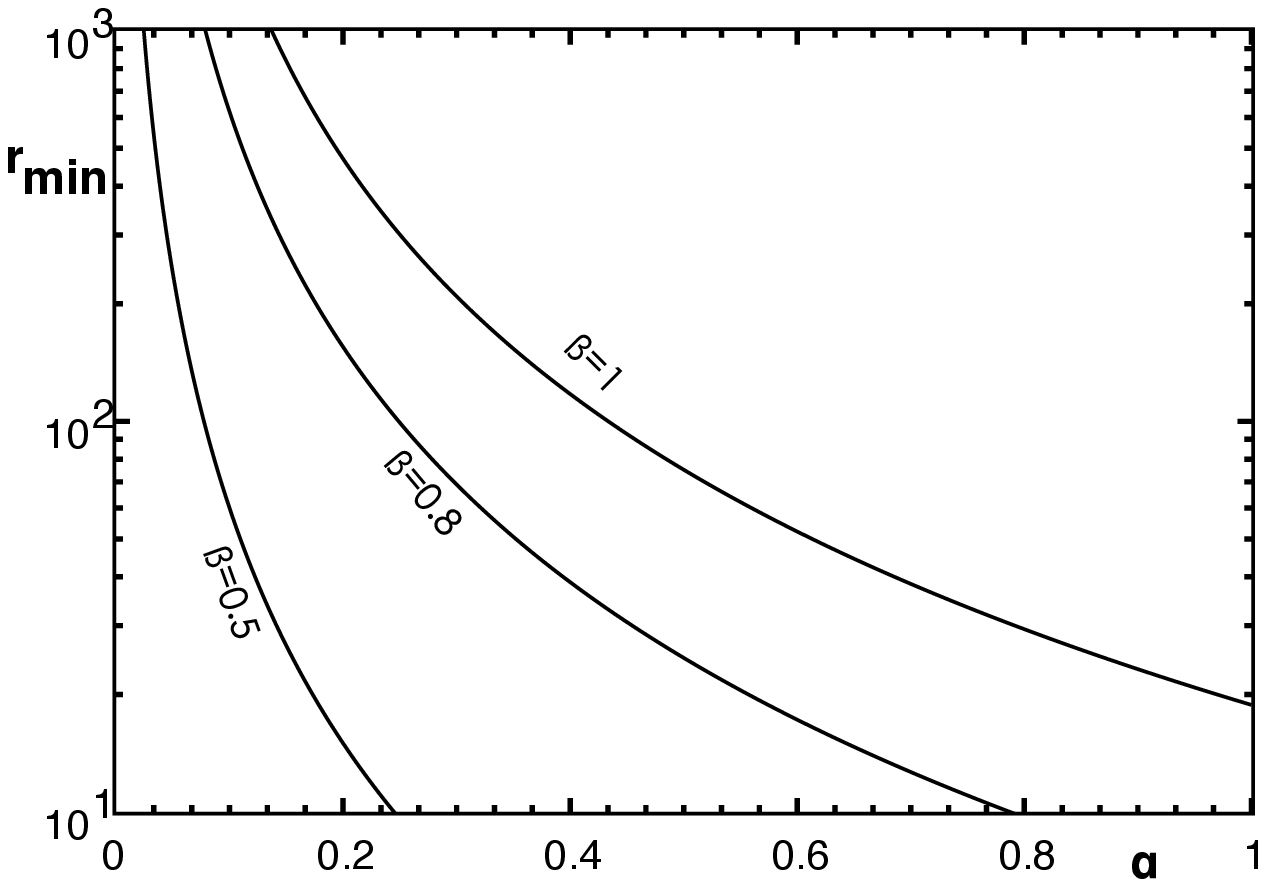}{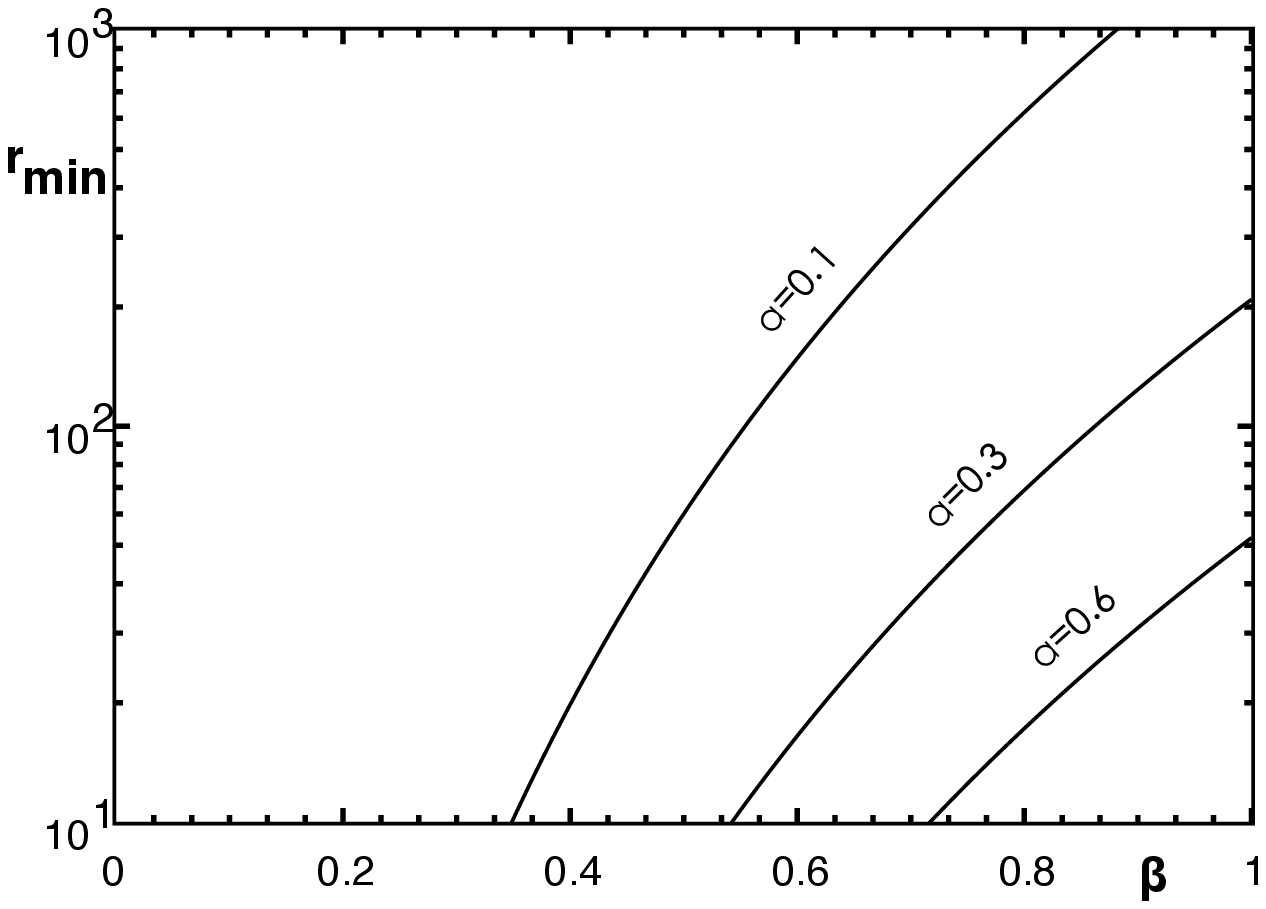}
\caption{\label{f:r_trs-alphabeta} The variation of the minimal truncation radius (in terms of the 
Schwarzshild radius) with respect to the 
viscosity parameter (left panel) and magnetic field parameter (right panel). An increase in $\alpha$ 
or/and a decrease in $\beta$ results in a decrease in the truncation radius.}
\end{figure}

An important feature of Eq.(\ref{mdot-max}) is that the transition accretion rate ($\dot m_{\rm max}$) 
strongly depends on the viscosity parameter. However, the influence of magnetic field on $\dot m_{\rm max}$ 
is limited to within a factor of 1.33 from zero-magnetic field ($\beta=1$) to an equipartition field 
($\beta=0.5$), as shown in Figure \ref{f:mdot_trs-alpha}. This indicates that the accretion rate 
characterizing the state transition is primarily determined by the viscosity.  On the other hand, the 
truncation radius is more sensitive to the magnetic field parameter than to the viscosity parameter. The 
increase in viscosity or/and magnetic field results in a decrease in the truncation radius before a transit 
to a soft spectral state, as shown in Figure \ref{f:r_trs-alphabeta}.  Hence, the observed transition 
luminosity (Eddington ratio) provides a constraint on the viscosity parameter and the observationally 
inferred truncation radius constrains the magnetic field parameter (see Eq.(\ref{tr-min})).  An 
approximate estimate of the value of the viscosity and magnetic field parameters in terms of the 
transition accretion rate ($\dot m_{\rm trs}$) and corresponding truncation radius ($t_{\rm trs}$) 
is 
\begin{eqnarray}
\alpha=0.20 \({\dot m_{\rm trs}\over 0.01}\)^{0.459} \({r_{\rm trs}\over 100}\)^{0.038} \label{alpha}
\approx  0.20 \({\dot m_{\rm trs}\over 0.01}\)^{0.459}\\ 
\beta=0.73 \({\dot m_{\rm trs}\over 0.01}\)^{0.185} \label{beta}\({r_{\rm trs}\over 100}\)^{0.216}.
\end{eqnarray}
These equations show that the viscosity parameter is determined by the transition accretion rate, which 
is little affected by the uncertainty in the truncation radius. On the other hand, the magnetic field 
parameter is constrained by both the transition accretion rate and corresponding innermost radius. Such 
results can be understood as follows.  A transition is triggered when the mass supply rate reaches the 
maximal mass evaporation rate. This maximal evaporation rate depends strongly on the viscosity parameter 
($\propto \alpha^{2.34}$) because  an increase in $\alpha$ leads to efficient heating in the inner corona. 
The effect of the magnetic field on the maximal evaporation rate is much weaker than the effect of 
viscosity due to a competition between its tendency to increase the evaporation as a result of energy 
balance and to decrease the evaporation as a result of pressure balance.  Thus, the transition accretion 
rate is mainly determined by the viscosity, providing a constraint on $\alpha$. However, the corresponding 
truncation radius depends both on $\beta$ and $\alpha$.

\begin{figure}

\plotone{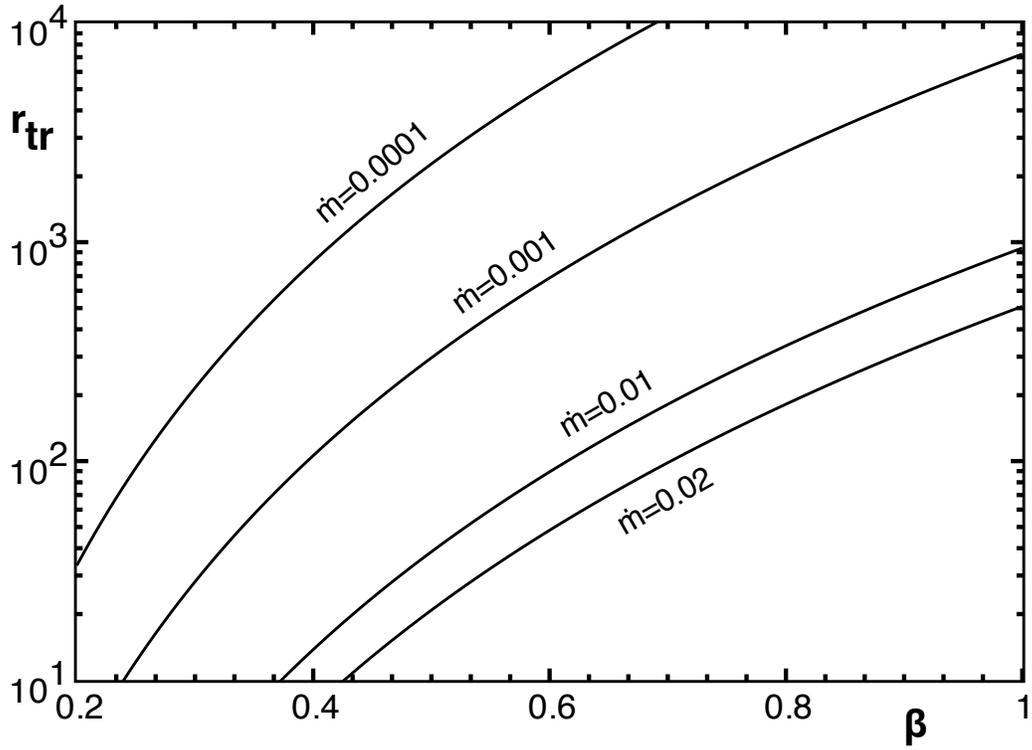}
\caption{\label{f:r_tr-beta} The variation of the truncation radius (in terms of the Schwarzschild radius) 
with respect to the magnetic field 
parameter $\beta$ in the low/hard state as predicted by the disk evaporation model. The truncation 
radius depends strongly on the magnetic field. The effect of the accretion rate is shown by different 
curves.  Since there is little influence from the viscosity parameter, $\alpha$ is fixed at 0.3.}
\end{figure}

Objects with very low accretion rates are far from the transition state. In this case, the disk is 
truncated at a large distance, $r_{\rm tr}>r_{\rm trs}$.  The truncation radius depends not only on 
the viscosity and magnetic field parameters, but also on the accretion rate. An approximate fit to 
the numerical data for $\alpha=0.3$ and $\beta=1$ yields an expression for the truncation radius 
given by 
\begin{equation}\label{tr-specific}
r_{\rm tr}\approx 15.9 \dot m^{-0.886}.
\end{equation}
This relation is extrapolated for different values of the viscosity and magnetic field parameters as 
(for details see Taam et al. 2012),
\begin{equation}\label{tr-radius}
r_{\rm tr}\approx 940\({\dot m\over 0.01}\)^{-0.886}\({\alpha\over 0.3}\)^{0.07}\beta^{4.61},
\end{equation}
which is only valid for accretion rates less than  half of the maximal accretion rate. 
The dependence of truncation radius on the magnetic field parameter and the accretion rate is 
shown  in Figure \ref{f:r_tr-beta}.  It can be seen from the figure and Eq.(\ref{tr-radius})
 that the truncation radius is strongly dependent on the magnetic field parameter, 
while it is only very weakly dependent on the viscosity parameter, unlike the minimal truncation radius 
at transition (Eq.\ref{tr-min}). That is, the viscous parameter little affects the truncation radius if 
$\alpha$ in the hot accretion flow is not significantly different from 0.3.  Therefore, the magnetic 
field parameter can also be constrained by the accretion rate and truncation radius from spectral fits 
to the observations of low/hard state objects by the expression 
\begin{equation}\label{beta-tr}
\beta\approx  \({\dot m\over 0.01}\)^{0.192}\({r_{\rm tr}\over 1000}\)^{0.217}. 
\end{equation} 

We note from Eqs.(\ref{alpha}) and (\ref{beta}) or (\ref{beta-tr}) that the viscosity and magnetic 
field parameters predicted by the disk evaporation model are in a range, i.e., $\alpha\sim 0.1-0.5$ 
and $\beta\sim 0.5-1$ for the typical accretion rate and truncation radius either immediately before 
transition or in the low hard state.  The values of $\alpha$ and $\beta$ do not vary steeply with 
accretion rate and truncation radius.  Therefore, we expect that the viscosity and magnetic field 
parameters can be approximately constrained by combining the model predictions and observational data, 
provided that the spectral state transition and disk truncation are determined (Meyer et al. 2000b; 
Liu et al. 2002; 2009; Taam et al. 2012).

\section{Constraints on the Viscosity and Magnetic Fields from Observations}
Given the mass of the black hole and the luminosity of objects near transition, the accretion 
rate ($\dot m_{\rm trs}$) can be determined. The viscosity parameter is estimated from the disk 
evaporation model; that is, $\alpha$ is calculated from Eq.\ref{alpha} where the truncation radius 
can be taken as $100 R_S$ as its value only very slightly affects $\alpha$. If this radius during 
transition can also be determined from the fitting of observed spectra, the magnetic field parameter 
can be estimated from Eq.\ref{beta}.  In this manner, both the viscosity and magnetic field parameters 
in objects at state transition are constrained.  For objects in a very low state, which are far from 
the transition state, the magnetic field parameter is estimated from Eq.\ref{beta-tr}, if the accretion 
rate and truncation radius can be determined from 
modeling the observed spectrum.  However,  the viscosity parameter for systems in a very low state can 
not be well constrained by the disk evaporation model since it does not affect the evaporation rate at 
large distances (see Eq.(\ref{tr-radius})).
 
\subsection{Constraint from Statistical Investigations}
Observationally, it is very difficult to detect the state transition of AGNs as the timescale for 
global accretion flow variability for 
supermassive black holes is much longer than for BHXRBs.  Thus, it is not possible to constrain 
the viscosity directly from the transition luminosity, as has been done in BHXRBs (Qiao \& Liu 2009). 
Nevertheless, evidence has been presented for a change in the accretion mode based on large-amplitude 
X-ray variability (Yuan et al. 2004) and the break of the X-ray photon index (see Constantin et al. 
2009).  The turning point in the relation between the photon index and the Eddington ratio occurs near 
$L/L_{\rm Edd} \approx 0.01$  (Constantin et al. 2009), which is similar to that exhibited by BHXRBs, 
provides empirical evidence for an intrinsic switch in the accretion mode. This yields an estimate 
for the averaged value of $\alpha$ at transition of $\sim 0.2$ as calculated from Eq.\ref{alpha}.

In a recent complementary investigation by Best \& Heckman (2012) a similar conclusion is deduced 
based on a large sample of radio-loud AGNs, showing that sources characterized by highly excited 
optical emission features typically have accretion rates between one per cent and 10 per cent of 
the Eddington rate, whereas low-excitation sources predominately accrete at rates below one per 
cent Eddington.  This implies a change of accretion mode taking place at an Eddington ratio of 0.01 
from a thin disk in the high-excitation sources to an ADAF in the low-excitation sources. The value 
of the viscosity parameter is also estimated to be $\alpha \sim 0.2$ in average based 
on the critical accretion rate of 0.01 for these radio loud AGNs.

\subsection{Constraint from Spectral Fits to AGN-powered LINERs}
Radiations in the optical, UV and X-ray bands of AGN are commonly thought to originate from 
a cold disk and hot accretion flow.  To fit the observed optical, UV, X-ray (and even radio) emissions of 
LLAGNs,  a truncated disk connected to an inner ADAF is often adopted.  The accretion rate and truncation 
radius of the disk are determined 
from spectral fits as shown by Quataert et al. (1999),  
Yuan \& Narayan (2004), Xu \& Cao (2009), and Nemmen et al. (2006; 2013). 
Specifically, the accretion rate and  truncation radius are taken as the main fitting parameters.  The 
overall continuum from radio to X-rays  are dominantly produced by the ADAF, where the mass flow 
 rate (expressed as a function of radius, $\dot m \propto r^s$ for $r < r_{\rm tr}$ ) determines the 
spectral shape.  The truncated disk can contribute to the optical/UV or infrared as its strength 
and peak frequency depend on the truncated radius and accretion rate. 
The fitting results from the literature have been compared with the disk evaporation model in 
detail in Taam et al. (2012). In 
Table \ref{Table1-AGN}, we list the accretion rate and truncation radius determined from the spectral 
fits to LINERs, and the values for 
the inferred viscosity and magnetic field parameters. It can be seen that the viscosity parameter is $\sim 
0.2 - 0.33$  and the magnetic field parameter is $0.5<\beta<1$. The inferred values of $\beta$ indicate 
that the magnetic pressure is less than or comparable to its equipartition value with gas pressure.   

\begin{table*}[t]
\caption{Viscosity parameter, $\alpha$, and magnetic field parameter, $\beta$, as constrained by 
accretion rates, $\dot m$, and truncation radii, $R_{in}$, in AGN-powered LINERs. 
Objects with relatively high accretion 
rates are assumed to be near transition so that both $\alpha$ and $\beta$ can be constrained, whereas 
objects with small accretion rates are regarded as in low/hard state and hence only $\beta$ can be 
constrained.}
\label{Table1-AGN}
\begin{tabular} {l|ccc|cc}\hline
Source & $R_{in}/R_{\rm S}$ & $\dot{m}$  & Reference & $\alpha$ & $\beta$\\\hline
 M81 & 100 & 0.01   & 1& 0.2& 0.73\\
  & 360 & 0.003   & 3& -& 0.65  \\ 
 NGC4579 & 100 & 0.02   & 1&0.33&0.89\\
 XMM J021822.3-050615.7 & 60 & $0.01$  & 2 &0.2&0.66\\
 NGC1097& 225 & $6.4\times 10^{-3}$ & 3&-&0.67 \\
 M87 & $10^4$ & $5.5\times 10^{-4}$ & 3 &-&1 \\
 NGC3398&$\ga500$&$10^{-3}$&3&-&$\ga$0.56\\
  NGC4278 & 100 & $4\times 10^{-3}$ & 3 &-&0.51\\
  \hline
\end{tabular}
\begin{list}{}{}
\item References. 1. Quataert et al. 1999; 2. Yuan \& Narayan  2004; 3. Nemmen et al. 2013
 \end{list}
\end{table*}

\subsection{Constraint from Spectral Fits to Simultaneous Optical-to-X-ray Observations of AGNs}
\begin{figure}

\plotone{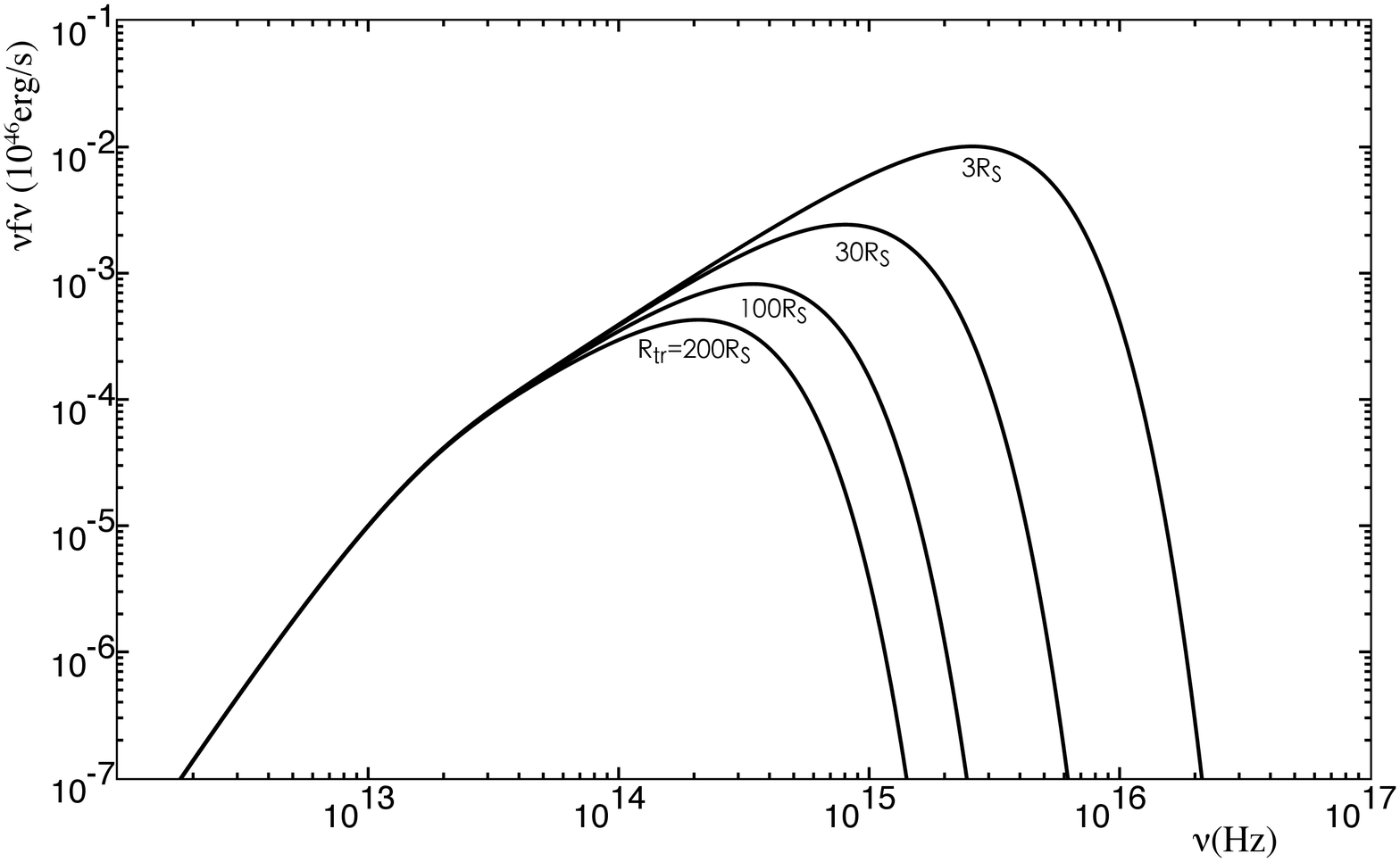}
\caption{\label{f:spectrum-tr} The multi-color blackbody spectrum for a typical LLAGN with black hole 
mass of $10^8M_\odot$ and accretion rate $\dot M=0.02 \dot M_{\rm Edd}$.  Curves are labeled 
corresponding to a truncation radius $R_{\rm tr}=3R_S, 30R_S,100R_S,200R_S$.  The disk emission extends 
to higher wavebands (i.e., optical or UV) with decreasing truncation radius.}
\end{figure}

Recently, Vasudevan \& Fabian (2009) and Vasudevan et al. (2009) modeled the simultaneous optical to 
X-ray emission of AGNs with a full disk (extending to the ISCO) plus a power law.  From the 
spectral fits over a wide wavelength coverage the bolometric luminosity and hard X-ray luminosity can 
both be determined, thus allowing a determination of the bolometric correction for observations based on 
hard X-rays alone, defined as the ratio of the bolometric luminosity to the 2-10keV luminosity.
For the low absorption AGN sample (Vasudevan et al. 2009) the bolometric corrections 
for the hard X-rays (2-10keV) are found to cluster within 
10-20 with the hard X-ray photon indices ranging from 1.5 to 2.  These features are in contrast to the 
properties of high luminosity AGNs (HLAGNs Vasudevan \& Fabian 2009) and are more similar to LLAGNs 
(Ho 2008). Combining the spectrum features with the low Eddington ratios (mostly a few percent or lower) for this 
sample,  we speculate that the objects are in an intermediate state, where the emission can originate 
from an ADAF surrounded by a corona and a disk truncated at a small radius. In this case, the optical 
radiation from these objects can be fit by a thermal spectrum produced by a multi-color blackbody from 
the truncated disk. This can be seen from the effective temperature of a truncated disk, which is 
expressed as, 
\begin{equation}\label{Teff-trun}
\begin{array}{ll}
T_{\rm eff}(R)=&6.237\times 10^5 \({R\over R_s}\)^{-3/4}\[{1-\({R_*\over R}\)^{1/2}}\]^{1/4}\\
&\times \({M\over 10^8M_\odot}\)^{-1/4}\({\dot M\over \dot M_{\rm Edd}}\)^{1/4}K. 
\end{array}
\end{equation}
$T_{\rm eff}$ is about 4300K for a $10^8M_\odot$ black hole accreting at the critical transition rate 
$\dot M=0.027\dot M_{\rm Edd}$ with a corresponding disk truncation at $\sim 200R_S$  (assuming 
$\alpha=0.3$ and $\beta=1$).  Taking into account the Wien's displacement (of 2.82), the truncated 
disk emission marginally extends to the optical waveband.  A decrease of the truncated radius, e.g. by 
magnetic field effects, results in an increasing disk temperature and radiation peaking at 
optical/UV frequencies. In Figure \ref{f:spectrum-tr}, we plot the disk spectrum for a 
typical LLAGN with $m=10^8$, $\dot m=0.02$ and disk truncation radius at $3R_S$, $30R_S$, $100R_S$ 
and $200R_S$ respectively. The figure clearly shows that the emission from the truncated disk extends 
to the optical and even to the UV band, with the flux peak and luminosity depending on the 
truncation radius.

Therefore, we expect that most objects in Vasudevan et al. (2009) could be in an intermediate state 
near the phase of transition between HLAGNs dominated by a geometrically thin disk and LLAGNs 
dominated by a small inner ADAF and a truncated disk.  If this is the case, we can constrain the 
viscosity and magnetic field parameters with the disk evaporation model.
 
In Table \ref{table2-AGN} we list objects with bolometric correction $k_{\rm 2-10keV}\la 20$, photon 
index $\Gamma\la2$  and Eddington ratio $L_{\rm bol}/L_{\rm Edd}<0.1$, taken from the sample of 
Vasudevan et al. (2009) and Vasudevan \& Fabian (2009).  Spectra for these objects are distinct from 
those with a high Eddington ratio and bolometric correction as shown in the spectral fits of Vasudevan 
\& Fabian (2009).  Observations for NGC 3227, NGC 3516 (Vasudevan \& Fabian 2009), and Mark590 (Vasudevan 
et al. 2009) are not included in the table because of their very low Eddington ratios, which implies 
that these objects probably are in the low/hard state rather than near the transition state.  
 
\begin{figure}
\plotone{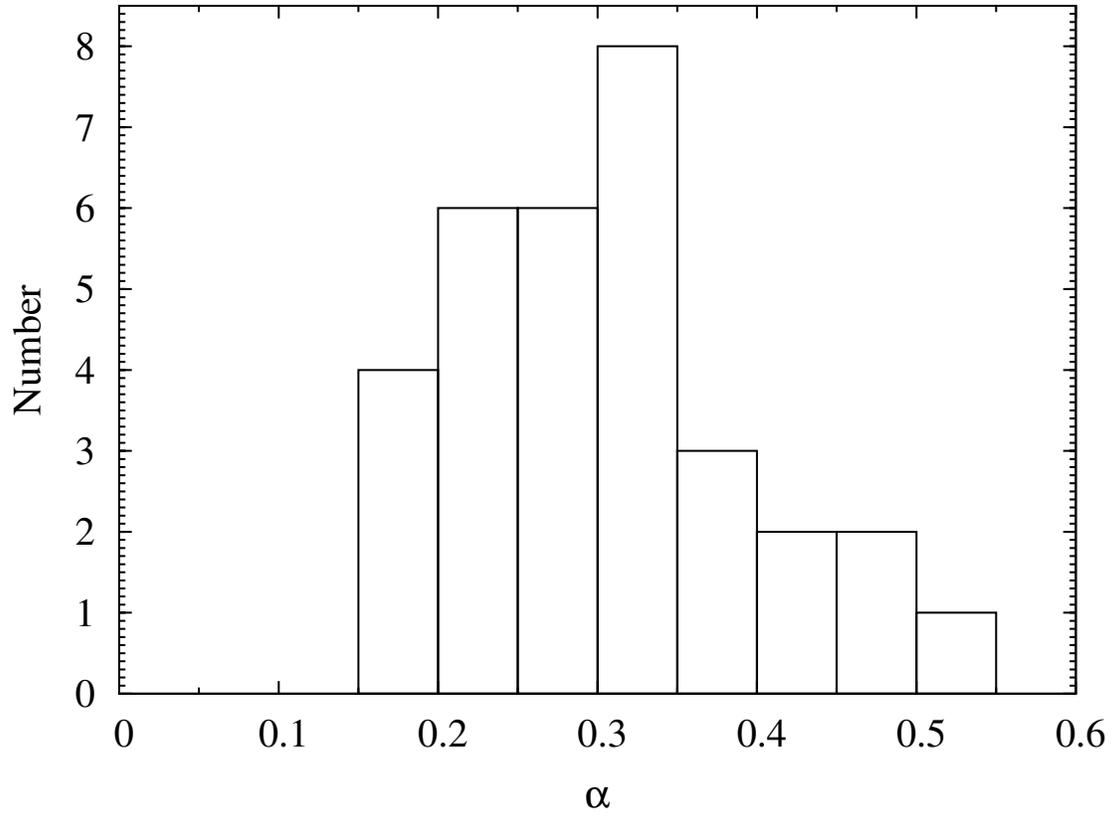}
\caption{\label{f:alpha} The distribution of the viscosity parameter of AGNs listed in Table \ref{Table1-AGN} and Table \ref{table2-AGN}.  It is shown that the value of viscosity parameter is clustered in 0.2-0.3.}
  
\end{figure}

We estimate the viscosity parameter for these objects by Eq.\ref{alpha} and list their values in the 
last column of Table \ref{table2-AGN}. The distribution of the value of $\alpha$ is plotted in Figure \ref{f:alpha}.   It can be seen that $\alpha$ mostly ranges from 0.2 to 0.35, 
which is only slightly affected by the (unknown) truncation radius. 

Within this framework,  the observed emission in the infrared or optical waveband indicates a disk 
temperature of 3000 to $10^4$K.  For black hole masses $\sim 10^8 M_\odot$ and accretion rates of 
a few percent of the Eddington rate, as for the objects listed in Table \ref{table2-AGN}, the disk 
should be truncated at distances of $\sim 100R_S$ or smaller to produce the IR or optical bump (see 
Figure \ref{f:spectrum-tr}). The truncation radius predicted by the disk evaporation model is larger 
than $200R_S$ when magnetic effects are neglected. For disk truncation at a distance 
smaller than $200R_S$, the effect of magnetic fields in the disk evaporation process is necessary.
 
We point out that the Eddington ratio, inferred from the spectral fits, is smaller when a truncated 
disk is used to model the optical observations than when modeled by a full disk (Vasudevan et al 2009; 
Vasudevan \& Fabian 2009).  This follows from the fact that there is no contribution from the inner 
region cut out from the disk, leading to a smaller disk luminosity (see Figure \ref{f:spectrum-tr}). 
This effect can be approximately neglected when the X-rays from the ADAF and the corona are the 
dominant component to the bolometric luminosity. With a small bolometric correction to the 2-10keV 
luminosity for objects listed in Table \ref{table2-AGN}, the Eddington ratios calculated from the 
spectral fits with a full disk+power law (Vasudevan \& Fabian 2009; Vasudevan et al. 2009) are a 
reasonable approximation to the intrinsic Eddington ratio even if the innermost disk is truncated.  
We note that a decrease in the Eddington ratio by a factor of five due, for example, to the absence 
of innermost disk, leads to only a decrease in the estimated viscosity parameter by a factor of two, 
for which $\alpha$ would be in a range of 0.1--0.25. Accurate spectral fits to the sources are 
reserved for a future study, but it will only slightly modify the value of the inferred 
viscosity parameter.
 
If the broad line region is associated with the disk through winds (e.g., Emmering et al. 1992), the truncation of the disk leads to the truncation of the broad line region (Liu \& Taam 2009).  
The presence of broad emission lines could be in conflict with a model where the inner disk is truncated. 
However, the disk is not truncated at a large distance ($R\la 200R_S$) for objects with relatively high 
accretion rates and relatively strong magnetic fields.  The emission lines can still occur, though they would 
not be expected to be very broad. This is not in contradiction with the fact that emission lines are 
observed in some of the objects listed in Table \ref{table2-AGN}.
  
\begin{table}
\caption{Viscosity parameter constrained from simultaneous optical-to-X-ray observations of AGNs }
\label{table2-AGN}      
\centering                          
 \begin{tabular}{l| llll|c}\hline\hline
 Source & $\Gamma$& $k_{2-10keV}$ & $L_{\rm bol}/L_{\rm Edd}$ &Ref. &$\alpha$\\\hline
 {\tiny 1RXS J045205.0+493248}	&1.86&12&0.067&1&0.48\\
{\tiny 2MASX J21140128+8204483}&1.85&8.1&0.025&1&0.30\\
3C 120 	&1.78&12.4&0.030	&1&0.33\\
3C 390.3(1)&1.74&6.33	&0.047&2&0.41\\
3C 390.3(2)&1.75&9.29&0.074&1&0.50\\
Ark 120	&1.90&17.79&0.028&1&0.32\\
ESO 490-G026	&1.91&12.1&0.022&1&0.29\\
ESO 548-G081	&2.03&13.06&0.015&1&0.24\\
Fairall 9	&1.81&10.5&0.019&2&0.27\\
IRAS 05589+2828&1.61&11.2	&0.008&1&0.18\\
MCG +04-22-042	&1.94&12.97&0.021&1&0.28\\
Mrk 1018	&1.95&12.07&0.027&1&0.31\\
Mrk 279	&1.88&9.83&0.007&1&0.17\\
Mrk 509	&1.83&14.84&0.019&1&0.27\\
Mrk 590	&1.88&8.8&0.01&2&0.20\\
Mrk 79 &1.91&10.5&0.031&2&0.34\\
Mrk 841	&1.89&16.9&0.021&1&0.28\\
NGC 3783(1)&1.53&7.02&0.043&2&0.39\\
NGC 3783(2)&1.50&8.0&0.036&2&0.36\\
NGC 4051&2.07&15.1&0.015&2&0.24\\
NGC 4151(1)&1.50&15.64&0.056&2&0.44\\
NGC 4151(2)&1.50&17.38&0.062&2&0.46\\
NGC 4593(1)	&1.87&7.7&0.037&2&0.36\\
NGC 4593(2)	& 1.62&9.89&0.009&1&0.19\\
NGC 5548(1)	&1.65&10.1&0.024&2&0.30\\
NGC 5548(2)	&1.51&8.8&0.009&1&0.19\\
NGC 7469	&1.98&14.33&0.010&1&0.20\\
NGC 985	&1.80&12.3&0.020&1&0.28\\
WKK 1263 	&1.68&23&0.032&1&0.34\\\hline 
 \end{tabular}
 \begin{list}{}{}
 \item Ref. 1. Vasudevan et al. 2009; 2. Vasudevan \& Fabian 2009
 \end{list}
 \end{table}  

\section{Discussion and Conclusion} 
The viscosity and magnetic field parameters in the accretion flow around black holes are estimated 
for low and intermediate luminosity AGNs.  According to the disk corona evaporation/condensation 
model, the accretion in low-accretion systems occurs via an inner ADAF connected to an outer disk as a 
consequence of the interaction between the disk and corona.  The truncation radius of the thin disk 
in the low/hard state is determined by the accretion rate, magnetic field and viscosity parameters.  A 
transition from an ADAF dominant accretion state to a full geometrically thin disk is triggered when 
the accretion rate reaches a critical value dependent on the viscosity parameter.  In the framework 
of this model, the viscosity parameter in the hot accretion flow is constrained from the observed 
transition luminosity, and the magnetic field parameter can be estimated from the spectral fits to 
the low hard states.  It is found that the viscosity parameter is $\sim 0.17 - 0.5$, but with values
clustered about 0.2-0.3.  Such values are consistent with those inferred from BHXRBs, which undergo 
outburst and are high compared to those deduced from hydrodynamical turbulence models (e.g. Kato 
1994) and from MHD simulations (for a review see King et al. 2007). The magnetic field parameter 
is found to range from 1 to 0.5, corresponding to disk truncation where magnetic field effects are 
unimportant to cases where magnetic pressure is in equipartition with the gas pressure. 
 
\subsection{Origin of Hard X-ray Emissions}
The hard X-ray emission in the low/hard state is assumed  to originate from the ADAF in this study. 
This is reasonable for objects with Eddington ratios higher than $10^{-6}-10^{-5}$ according to the 
prediction of Yuan \& Cui (2005). Observational investigations of AGN-powered LINERs support an ADAF 
origin, however, the contribution of a jet based on the fits to the fundamental plane (Younes et al. 
2012) cannot be excluded.  In addition, spectral fits to the LLAGN (Nemmen et al. 2012) also show 
that both the jet and ADAF can fit the spectrum with different parameters for most of the objects 
in their sample.  

A high-spatial resolution study of the spectral energy distribution (SED) of the nearest LLAGNs 
(Fern\'andez-Ontiveros et al. 2012) shows a large diversity in the SED shapes in the LLAGN sample, 
some of which are very well described by the self-absorbed synchrotron process, while others present 
a thermal-like bump at $\sim 1\mu$.  The SEDs in the sample intrinsically differ from the SEDs of bright 
AGNs, suggesting that the inner accretion flow of AGNs undergoes changes with the decrease of 
the mass accretion rate, probably from a thin accretion disk to an ADAF.
 
\subsection{The Compton Effect}
Compton scattering can become especially important for a strong external seed photon field, 
contributed by the central accretion flow and local underlying disk.  In the low/hard states as considered 
here, the radiation from the inner ADAF is inefficient. Radiations from the outer disk are important for Compton cooling only at accretion rates close to the maximal evaporation rate. This can  cause a decrease in the maximal evaporation rate by a factor of   $\sim1/2$, leading to an underestimate of the viscosity parameter by a factor of $\sim 1.3$ for objects near transition. If a magnetic field 
is taken into account, it leads to a decrease of the evaporation rate at a given distance. As 
a consequence, the coronal density decreases and the disk radiation increases as more mass 
remains in the disk. The net result is that the Compton effect for disks with a magnetic field would 
be similar to the case without a magnetic field, as estimated above.  Nevertheless, since 
the disk is truncated at a smaller region before state transition, the Compton scattering could 
be important at small distances. We plan to investigate this possibility in the future. 

The existence of a geometrically thin disk at the transition from the soft to hard state, in contrast
to the transition from the hard to soft state, results in strong Compton cooling of the corona. This leads 
to a lower transition luminosity compared to that from the hard to soft state transition.
Such an effect has been interpreted as 
due to a hysteresis effect in the state transitions observed in the outbursts of BHXRBs (Meyer-Hofmeister 
et al. 2005; Liu et al. 2005). If the analogy of the accretion process in stellar mass black holes to 
super massive black holes extends to this phenomena as well, the AGNs with intermediate luminosities 
(corresponding to accretion rates $\sim 0.006-0.03$ can be either in a soft state approaching a 
transition to a hard state, or in a hard state evolving towards a soft state.  An object evolving from a 
disk dominant state could lead to an underestimate of the parameters by up to a factor of two. 
This could be the case for some of the objects listed in Table \ref{table2-AGN}. That is, an object 
evolves close to disk truncation, with most of the disk gas evaporated into the corona, leading to a 
weak un-truncated disk and relatively strong corona.  

A recent investigation on the local-radio AGN populations shows that the distribution of 
Eddington-scaled accretion rates in the high excitation radio galaxies (HERG) is distinctly higher, 
on average, than in the low excitation radio galaxies (LERG), supporting the scenario of thin disk 
dominant accretion in HERG and an ADAF dominant accretion in LERG (Best \& Heckman 2012). The overlap 
region in Eddington ratio for LERG and HERG may be analogous to the intermediate state observed in 
BHXRBs.  Here, systems evolving from a soft state to this intermediate state could exhibit high 
excitation lines with relatively weak radio emission, whereas systems evolving from the hard state 
may be characterized as low excitation radio loud sources.  The overlap region corresponding to an 
Eddington ratio in the range of 0.001 to 0.03, as shown in Fig.6 of Best \& Heckman (2012), suggests 
an intriguing connection within the framework of such a model. 

\acknowledgments
We are grateful to Weimin Yuan for discussion and comments on the manuscript. 
Financial support for this work is provided by the National Basic Research Program of China-973 Program 
2009CB824800 and by the National Natural Science Foundation of China (grants 11033007, 11173029 and U1231203). In 
addition, R.E.T. acknowledges support from the Theoretical Institute for Advanced Research in Astrophysics 
in the Academia Sinica Institute of Astronomy \& Astrophysics.


\end{document}